\documentstyle[twoside,fleqn,espcrc2,epsfig]{article}
\input{epsf.tex}

\title{A disorder parameter for dual superconductivity in gauge theories.}
\author{ \underline{A. Di Giacomo}, B. Lucini, L. Montesi, G. Paffuti\\
Dipartimento di Fisica and INFN, 2 Piazza Torricelli 56100 Pisa, Italy}

\begin{document}
\begin{abstract}
Dual superconductivity in the confining phase of gauge theories is discussed in
terms of a disorder parameter which vanishes in normal phase and is different
from zero in the superconducting phase.
\end{abstract}
\maketitle
\section{Introduction and conclusions.} There exists evidence from lattice
simulations that QCD vacuum is a dual superconductor in the confining phase,
and undergoes a transition to normal state at deconfinement
temperature\cite{1,2}.

This support the idea that confinement is produced by dual Meissner
effect\cite{3,4}.

However there are aspects of confinement which are not understood in this
scenario.

The evidence for dual superconductivity from lattice simulations can be put
into two categories: phenomenological  and  direct.

Phenomenological evidence is the observation of basic features which are
consistent with the picture:\\
1) Existence of string tension, as detected by the
area law behaviour of Wilson loops\cite{5}, indicating that confinement really
takes place in QCD.\\
2) Existence of flux tube configurations between static
$Q\bar Q$ pairs\cite{6,7}.\\
3) String like behaviour of flux tubes\cite{8}.

Direct evidence is instead  produced by  detection of monopole
condensation. A non zero vacuum expectation value for any operator carrying non
zero magnetic charge signals spontaneous breaking of magnetic $U(1)$ and hence
dual superconductivity\cite{9,1}. An alternative way of investigation
consists in detecting persistent (London) electric currents in the vacuum, which
are a direct consequence of dual superconductivity\cite{2}.

Monopoles which are expected to condense in the ground state to produce dual
superconductivity are $U(1)$ monopoles defined by a procedure known as
``abelian projection'', in terms of any local operator $\vec \Phi(x)$ belonging
to the adjoint representation of the gauge group. These Dirac monopoles are
located in the field configurations at the sites where $\vec \Phi(x)$ has
zeros. The equation $\vec \Phi(x) = 0$ identifies the world line of a monopole, we are referring to $SU(2)$ gauge group for simplicity of notation.

Thus there exist a functional infinity of abelian projections, and each of them
defines monopoles. The question is then:\\
Is any of these abelian projection
better than the others? or\\
Can we identify ``The abelian projection'', which
describes confinement?

An argument of continuity suggests that this is not the case: infinitesimal
changes of $\vec \Phi$ cannot produce dramatic effects as going from
condensation to absence of it. Moreover if one single abelian projection is at
work, the string tension in the adjoint representation is zero: one gluon in
$SU(2)$ (two of them in $SU(3)$) have zero $U(1)$ charge and there is no flux tube
between them.

The colour content of flux tubes should be in the direction of abelian
projected $U(1)$, since it is the abelian projected chromoelectric field which
is channeled into Abrikosov flux tubes.
A test on lattice shows that it is isotropically oriented in colour
space\cite{10,7}.

Our strategy to this problem is the following: we have a reliable disorder
parameter which can detect dual superconductivity\cite{9}. By use of it we
investigate in different abelian projections 1) The existence of dual
superconductivity in conjunction with confinement. 2) The type of
supeconductivity.

This systematic investigation is on the way. Our preliminary results are that
different abelian projection (Polyakov line, Field strength, max abelian)
identify monopoles which condense in the confined phase, and do not in the
deconfined one.

This agrees with the idea of t'Hooft that all abelian projections are
physically equivalent\cite{11}.

From the theoretical point of view all this indicates that QCD vacuum is more than a
$U(1)$ dual superconductor: some different, non abelian yet unknown mechanism is at
work, which shows up as $U(1)$ dual superconductivity in different abelian
projections.
\section{A few details about the disorder parameter.}
We define an operator $\mu(\vec x,t)$ which has magnetic charge $Q\neq0$. 
We  then
measure its correlation function with the charge conjugate operator $\bar\mu$; e.g.
\begin{equation}
\hskip-24pt
{\cal D}(t) = \langle \bar\mu(\vec 0,t),\mu(\vec 0,0)\rangle
\mathop\simeq_{t\to\infty} A \exp(-M t) + \langle \mu\rangle^2
\label{eq:1}\end{equation}
$M$ is the lowest mass in the sector with magnetic charge $Q$: $\langle\mu\rangle\neq 0$ signals
spontaneous breaking of magnetic $U(1)$ and hence dual superconductivity. We
also measure the penetration depth of the electric field in the vacuum $E_\Vert
= E(0) \exp(- m x)$. If $M \geq \sqrt{2} m$ the superconductor is type II\cite{12}.

${\cal D}(t)$ is the ratio of two partition functions\cite{9}
\begin{equation}
{\cal D} = \frac{\displaystyle Z[S + \Delta S]}{\displaystyle Z[S]}
\label{eq:2}\end{equation}
$\Delta S$ is the modification of the plaquettes $\Pi^{0i}$ in the action at
time 0, when the monopole is created and at time $t$, when it is destroyed. For
$U(1)$ the modification consists in the substitution
\begin{equation}
\hskip-20pt \Pi^{0i} = {\rm e}^{
i\theta^{0i}(x)}\Rightarrow
{\rm e}^{i(\theta^{0i}(x) - b^i(\vec x))}\qquad x^0 = 0
\label{eq:3}\end{equation}
\begin{equation}
\hskip-20pt \Pi^{0i} = {\rm e}^{
i\theta^{0i}(x)}\Rightarrow
{\rm e}^{i(\theta^{0i}(x) + b^i(\vec x))}\qquad x^0 = t
\label{eq:4}\end{equation}
$\frac{1}{e}\vec b(\vec x)$ being the vector potential produced by a Dirac
monopole sitting at the origin
\begin{equation}
\vec b(\vec x) = \frac{1}{2}\frac{\displaystyle \vec n\wedge\vec x}
{\displaystyle |\vec x\,|(|\vec x| - \vec n\cdot\vec x)}
\label{eq:5}\end{equation}
In computing $Z[S + \Delta S]$ the shift (\ref{eq:3}) can be reabsorbed in a
shift of the angle $\theta_i(\vec x,1)$, at $t=1$, which 
leaves the measure of the integral invariant. This produces at time 1
the following change
\begin{equation}
\theta_{ij}(\vec x,1)\to \theta_{ij}(\vec x,1) +
\Delta_i b_j - \Delta_j b_i\label{eq:6}\end{equation}
meaning that a monopole has been created. At the same time $\theta_{0i}(\vec
x,1)\to \theta_{0i}(\vec x,1) - b_i(\vec x)$ and again a change of variables
$\theta_i(\vec x,2)\to \theta_i(\vec x,2) - b_i(\vec x)$ exposes a monopole at
$t=2$, and a shift in $\theta_{0i}(\vec x,3)$ and so on, till $x^0 = t$ is
reached and $-b_i$ cancels with $+b_i$ in eq.(\ref{eq:4}).

In fact a more convenient quantity than ${\cal D}$ is $\rho = d \ln {\cal D}/d\beta$:
for
$\beta < \beta_c$, as $V\to \infty$, $\rho$ tends to a finite value, so that
$\mu = \exp(\int_0^\beta \rho(x)\,d x)$ for $\beta < \beta_c$ tends to a
positive function\cite{9}. For $\beta > \beta_c$ $\rho\sim - C\, V^{1/4}$ so that
$\mu\to 0$ in the infinite volume limit. 
Around $\beta_c$ a finite size scaling analysis allows a precise determination of
$\beta_c$, of the critical index by which $\mu\to 0$ at $\beta_c$, and of the
critical index of the correlation length. 

A typical behaviour of $\rho$ for $U(1)$ is shown in fig.1. The phase transition
appears as a sharp negative peak at $\beta_c$.

\par\noindent
\begin{minipage}{0.9\textwidth}
\epsfxsize = 0.5\textwidth
\epsfbox{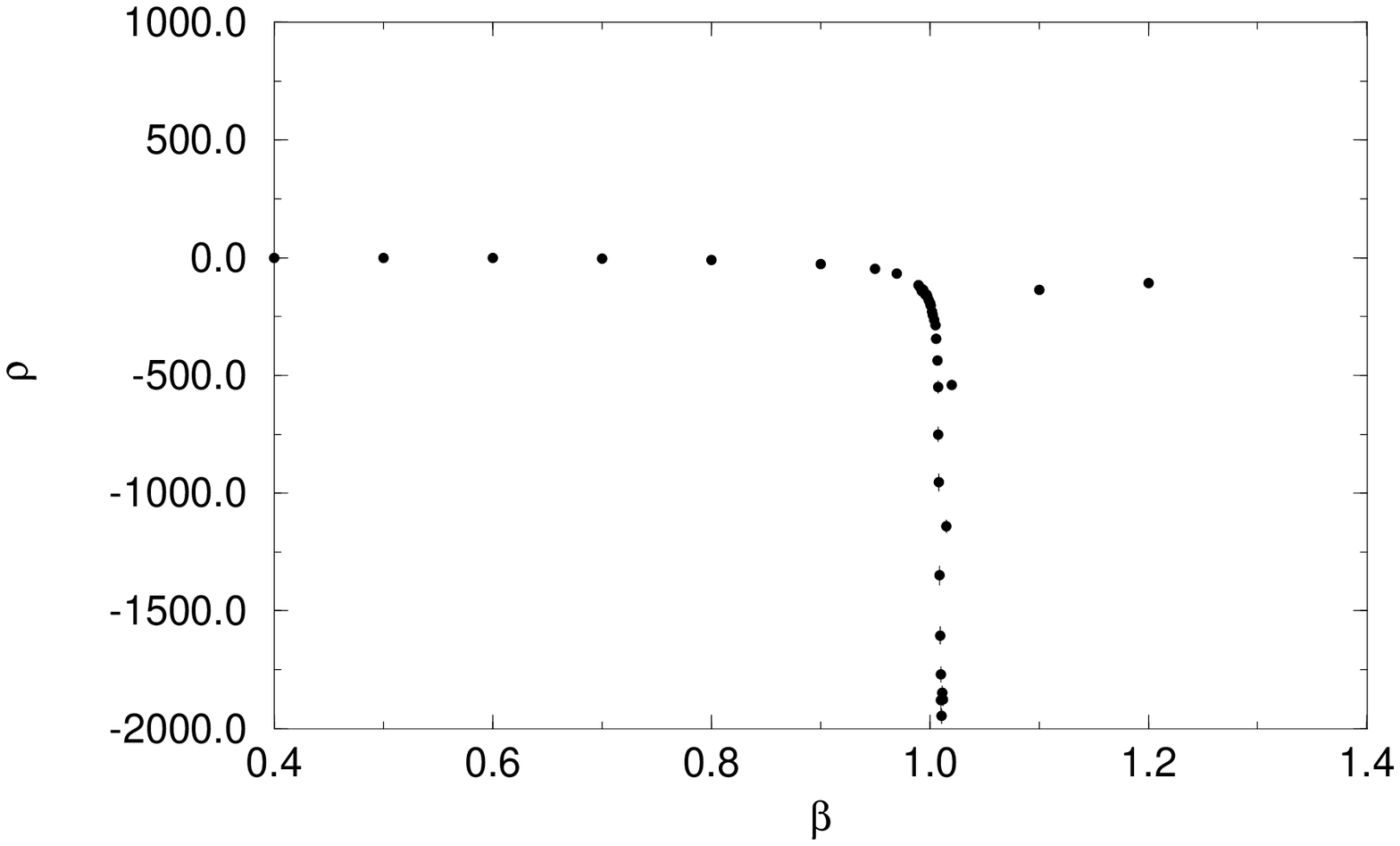}
\begin{itemize}\item[Fig.1]
 $\rho$ vs $\beta$ for  U(1).
The peak\\ is at the deconfining transition.
\end{itemize}
\end{minipage}
\vskip0.1in
A similar construction can be made for $SU(3)$ and $SU(2)$ gauge theories at finite
$T$, (fig.2,3). 
\vskip0.1in
\par\noindent
\begin{minipage}{0.9\textwidth}
\epsfxsize = 0.5\textwidth
\epsfbox{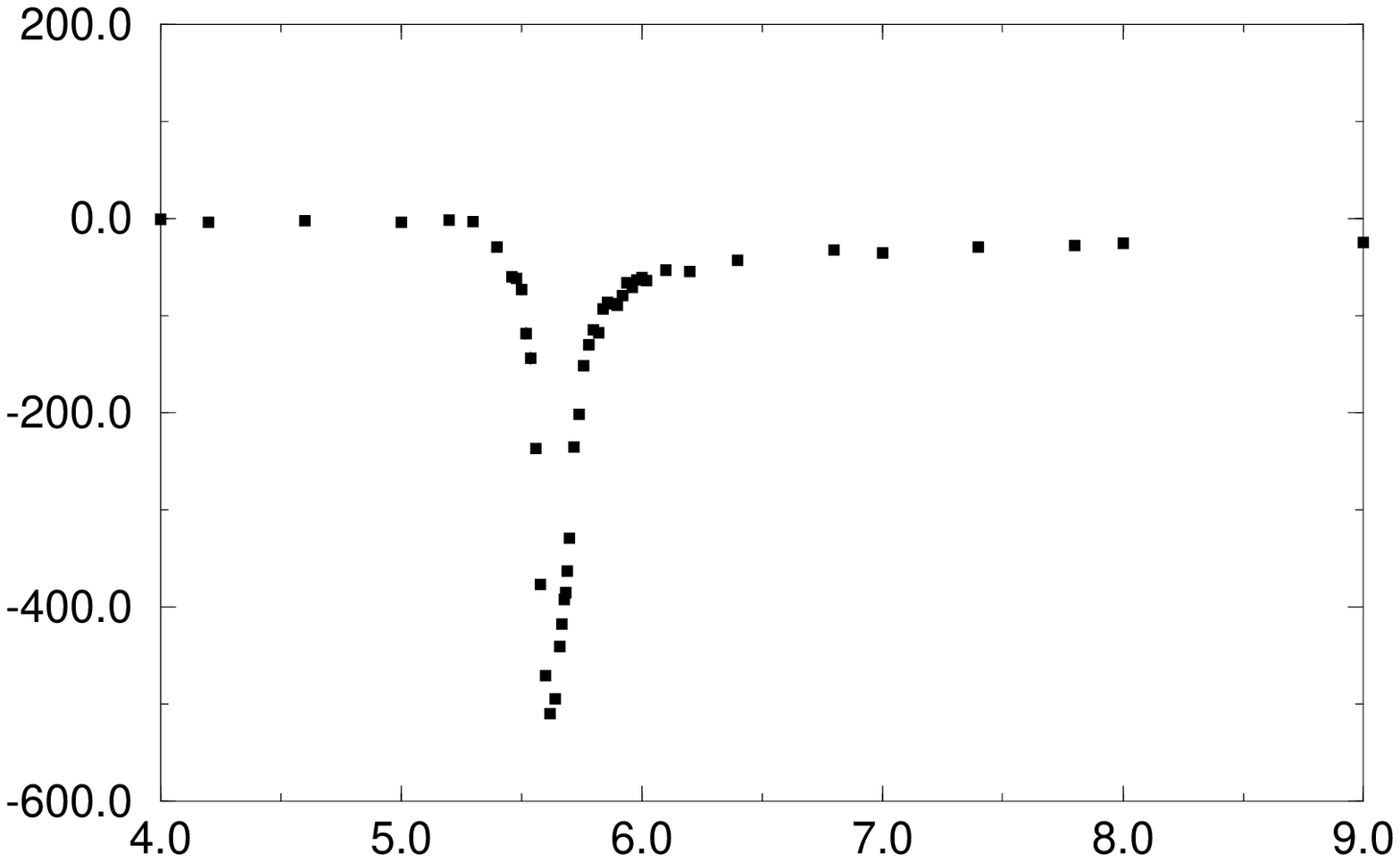}
\begin{itemize}\item[Fig.2]
$\rho$ vs $\beta$ for SU(3) at finite $T$.
The peak \\ is at the deconfining transition.
\end{itemize}
\end{minipage}
\vskip0.1in
\par\noindent
\begin{minipage}{0.9\textwidth}
\epsfxsize = 0.5\textwidth
\epsfbox{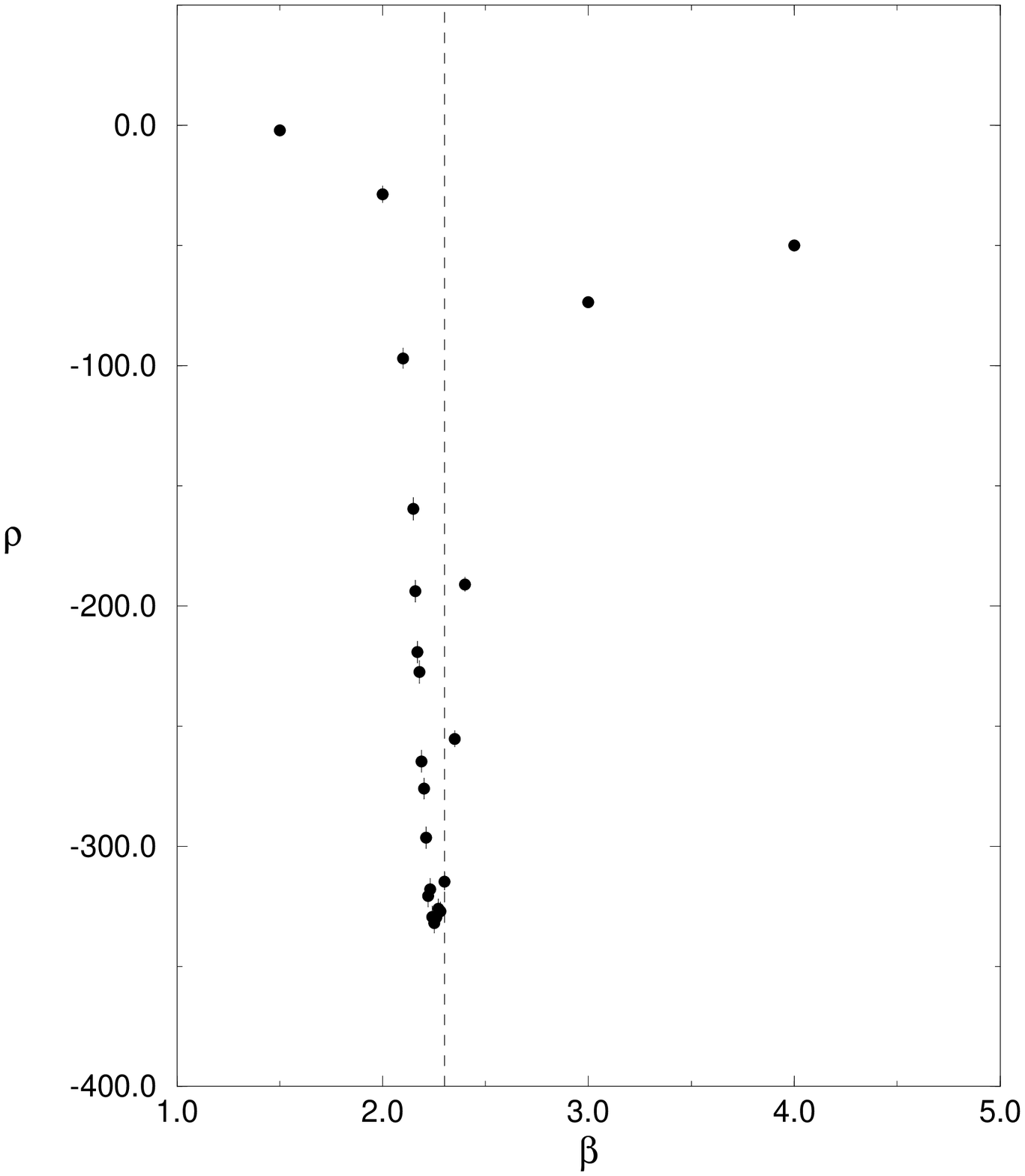}
\begin{itemize}\item[Fig.3]
$\rho$ vs $\beta$ for SU(2) at finite $T$.
The peak\\ is at the deconfining transition. 
\end{itemize}
\end{minipage}

More generally the construction works for phase transitions
produced by condensation of topological solitons in the vacuum: vortices in 3d $XY$ model\cite{13}, 
fig.4, $O(3)$ solitons in 3d Heisenberg model\cite{14}.

\par\noindent
\begin{minipage}{0.9\textwidth}
{\rotatebox{-90}{
\epsfxsize = 0.35\textwidth
\epsfbox{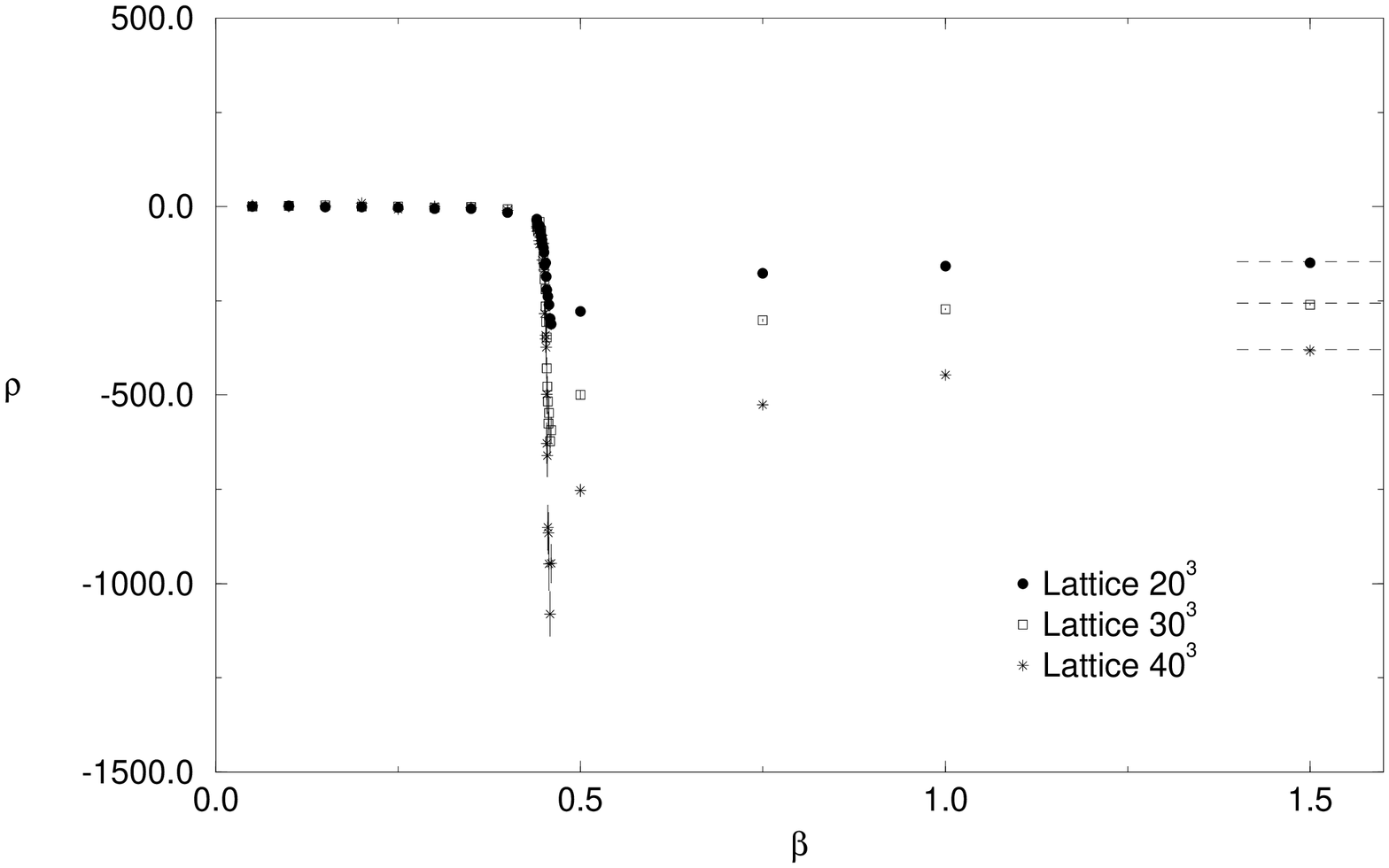}
}
}\par\noindent
\begin{itemize}\item[Fig.4]
$\rho$ vs $\beta$ for $X-Y$ model.
The peak \\ is at the transition to superfluid.
\end{itemize}
\end{minipage}
\vskip0.2in

In conclusion we have a reliable tool to investigate $U(1)$ dual
superconductivity. We are using it to explore abelian projected $U(1)$ in non
abelian gauge theories, in different abelian projections.


\begin{thebibliography}{999}
\bibitem{1}
L.Del Debbio, A.Di Giacomo, G.Paffuti and P.Pieri:
{\it Phys. Lett.}{\bf B 355} (1995) 255.
\bibitem{2} V. Singh,R.W. Haymaker, D.A. Brown: 
{\em Phys. Rev.} {\bf D47} (1993) 1715.
\bibitem{3}G. 't~Hooft, in ``High Energy Physics'', EPS
International Conference, Palermo 1975, ed. A.~Zichichi, Bologna 1976, p 1225.
\bibitem{4}S. Mandelstam: {\em Phys. Rep.} {\bf 23C} (1976) 245.
\bibitem{5} M. Creutz: {\em Phys. Rev.} {\bf D21} (1980) 2308.
\bibitem{6} R.W. Haymaker, J. Woziek: 
{\em Phys. Rev.} {\bf D36} (1987) 3297.
\bibitem{7}
A. Di Giacomo, M.~Maggiore and \v{S}.~Olejn\'{\i}k:
{\em Nucl. Phys.} {\bf B347} (1990) 441.
\bibitem{8} M.Caselle, R.Fiore, F.Gliozzi, M.Hasenbusch, P.Provero: 
{\em Nucl. Phys.} {\bf B486} (1997) 245.
\bibitem{9} A.Di Giacomo, G.Paffuti: {\em Detecting dual superconductivity in
gauge theory vacuum}, heplat 9707003.
\bibitem{10}J.Greensite, J.Winchester:
{\em Phys. Rev.} {\bf D40} (1989) 4167.
\bibitem{11}G. 't~Hooft, {\em Nucl. Phys.} {\bf B190} (1981) 455.
\bibitem{12} A.B. Abrikosov {\em JETP} {\bf 5} (1957) 1174.
\bibitem{13}G. Di Cecio,
A.Di Giacomo, G.Paffuti, M.Trigiante:
{\em Nucl. Phys.} {\bf B 489} (1997) 739.
\bibitem{14}
A.Di Giacomo, D. Martelli,G.Paffuti: in preparation.
\end{thebibliography}
\end{document}